\documentclass[prx, nofootinbib, twocolumn, amsmath, amssymb, superscriptaddress]{revtex4-2}

\usepackage{bbold}
\usepackage{hyperref}
\hypersetup{colorlinks=true, citecolor=blue, linkcolor=blue, urlcolor=gray}
\usepackage{graphicx}
\usepackage{bm}
\usepackage{color}
\usepackage{url}
\usepackage{tikz}
\usetikzlibrary{calc}
\usepackage{pgfplots}
\usepackage{amsmath,amssymb}
\usepackage{verbatim}
\usepackage{physics}
\usepackage{xcolor}
\usepackage{mwe}
\usepackage{tikz}
\usetikzlibrary{calc}
\usepackage{pgfplots}

\newcommand{\bea}{\begin{eqnarray}}
\newcommand{\eea}{\end{eqnarray}}

\newcommand{\td}{d}

\newcommand{\be}{\begin{equation}}
\newcommand{\ee}{\end{equation}}

\newcommand{\beal}{\begin{align}}
\newcommand{\eeal}{\end{align}}

\newcommand{\vn}{\vec{n}}
\newcommand{\cl}{\mathcal{L}}

\newcommand{\cz}{\mathcal{Z}}
\newcommand{\cd}{\mathcal{D}}
\newcommand{\co}{\mathcal{O}}

\newcommand{\rt}{\rho_{\text{top}}}

\begin{document}


\title{Topological dipoles of quantum  skyrmions} 

\author{Sopheak Sorn}
\email{sopheak.sorn@kit.edu}
\affiliation{Institute of Theoretical Solid State Physics, Karlsruhe Institute of Technology, 76131 Karlsruhe, Germany
}%
\affiliation{
Institute for Quantum Materials and Technology, Karlsruhe Institute of Technology, 
76131 Karlsruhe, Germany
}%
\author{Jörg Schmalian}
\affiliation{
Institute for Quantum Materials and Technology, Karlsruhe Institute of Technology, 
76131 Karlsruhe, Germany
}%
\affiliation{%
Institute of Theoretical Condensed Matter Physics, Karlsruhe Institute of Technology, 76131 Karlsruhe, Germany
}%
\author{Markus Garst}
\email{markus.garst@kit.edu}
\affiliation{Institute of Theoretical Solid State Physics, Karlsruhe Institute of Technology, 76131 Karlsruhe, Germany
}%
\affiliation{
Institute for Quantum Materials and Technology, Karlsruhe Institute of Technology, 
76131 Karlsruhe, Germany
}%

\date{\today}

\begin{abstract}
Magnetic skyrmions are spatially localized whirls of spin moments in two dimension, featuring a nontrivial topological charge and a well-defined topological charge density. We demonstrate that the quantum dynamics of magnetic skyrmions is governed by a dipole conservation law associated with the topological charge, akin to that in fracton theories of excitations with constrained mobility. The dipole conservation law enables a natural definition of the collective coordinate to specify the skymion's position, which ultimately leads to a greatly simplified equation of motion in the form of the Thiele equation. In this formulation, the skyrmion mass, whose existence is often debated, actually vanishes. As a result, an isolated skyrmion is intrinsically pinned to be immobile and cannot move at a constant velocity.
In a spin-wave theory, we show that such dynamics corresponds to a precise cancellation between a highly nontrivial motion of the quasi-classical skyrmion spin texture and a cloud of quantum fluctuations in the form of spin waves. Given this quenched kinetic energy of quantum skyrmions, we identify close analogies to the bosonic quantum Hall problem. 
In particular, the topological charge density is shown to obey the Girvin-MacDonald-Platzman algebra that  describes  neutral modes of the lowest Landau level  in the fractional quantum Hall problem. 
Consequently, the conservation of the topological dipole suggests that magnetic skyrmion materials offer a promising platform for exploring fractonic phenomena with close analogies to fractional quantum Hall states.
\end{abstract}


\maketitle

\section{Introduction}

\begin{figure}[t]
\centering
\includegraphics[width = 0.45\textwidth]{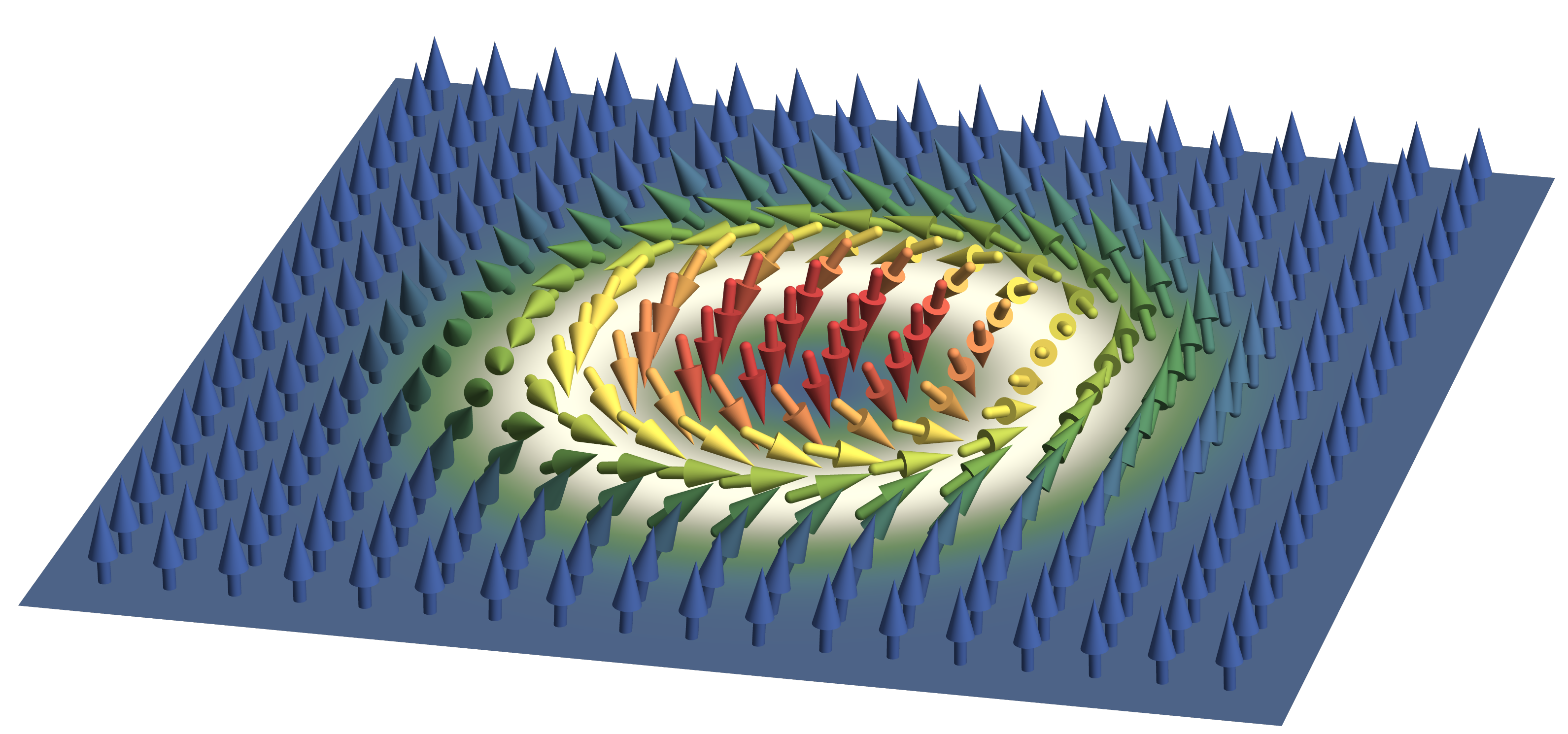}
\caption{Example of a topological skyrmion texture with charge $Q_{\rm top} =  - 1$. The arrows represent  the field $\vec n$ and the brightness of the background color indicates the magnitude of the associated topological density $\rt$. The density $\rt$ is only finite in  regions with spatially varying $\vec n$.}
\label{fig:skyrmion}
\end{figure}

Magnetic skyrmions are topological spin textures that have been studied intensively over the past decade and that allow for an unprecedented amount of individual control; for reviews see Refs.~\cite{Nagaosa2013, Wiesendanger2016, Jiang2017, Fert2017, EverschorSitte2018, Back2020, Goebel2021,petrovic2024}.
They can be created and investigated in a controlled manner not only in specifically designed magnetic heterostructures but also in a variety of bulk magnetic materials. The nontrivial topology endows skyrmions with a topological protection against small perturbations, which, alongside with their small size, drives a number of proposals to exploit skyrmions as novel bits in future information and data-storage technology \cite{Finocchio2021}. 
Moreover, skyrmions can imprint their topology on the conduction electrons in itinerant magnets, resulting in an emergent electric and magnetic field via a real-space Berry-phase mechanism \cite{Volovik1987,Ye1999,Bruno2004}. These emergent fields affect transport properties, giving rise to a topological Hall effect \cite{Neubauer2009, Lee2009} as well as a topological Nernst effect \cite{Shiomi2013, Hirschberger2020}. Vice versa, a current of electrons affects the dynamics of skyrmions, yielding a skyrmion Hall effect \cite{Thiele1973}. In addition, magnetic skyrmions possess characteristic magnon eigenmodes \cite{Mochizuki2012}, and their non-trivial topology is also reflected in magnon Chern bands of skyrmion crystals \cite{Roldan2016,Garst2017}.
Skyrmions are amenable to imaging techniques, such as transmission electron microscopy, scanning tunneling microscopy or scanning transmission X-ray microscopy, allowing to determine the position of magnetic skyrmions, their internal spin arrangement, and, if combined with the proper time resolution, also the skyrmion dynamics \cite{petrovic2024}. 

The non-trivial topology of magnetic skyrmions is characterized by the topological charge  
\bea \label{eq:TopCharge}
Q_{\rm top}(t) = \int_{V} \td \vec{x} \ \rt(\vec x, t),
\eea
where the integral is over the two-dimensional plane with area $V$, and the topological density is defined by 
\begin{align}
\label{eq:TopDensity}
\rt &= \frac{1}{4\pi} \vn \cdot (\partial_1 \vn \times \partial_2 \vn).
\end{align}
The local magnetization is specified by a three-component unit vector field, $\vn(\vec{x}, t)$, as a function of the spatial position  $\vec x = (x_i)$ with $i = 1,2$ and time $t$.
For boundary conditions where $\vec n(\vec x, t)|_{\partial V}$ is constant on the surface $\partial V$ of the two-dimensional volume, the charge $Q_{\rm top}$ is a time-independent integer number. An example of a magnetic skyrmion texture with $Q_{\rm top} = -1$ is shown in Fig.~\ref{fig:skyrmion}. 
At the focus of the present work is the first moment of topological charge,
\bea \label{eq:TopDipole}
D_i(t) = \int_{V} \td \vec{x} \ x_i\, \rt(\vec x, t),
\eea
i.e., the dipole moment of the topological charge distribution. Assuming that the dynamics of the magnetization $\vec n$ is governed by the Landau-Lifshitz equation, it was shown in a seminal work by Papanicolaou and Tomaras \cite{Papanicolaou1991} that the dipole moment $D_i$ is related to the linear momentum of the texture, $P_i = 4\pi s \epsilon_{ki} D_k$, where $s$ is the spin density and $\epsilon_{12} = -\epsilon_{21} = 1$. For a system with translation invariance the conservation of linear momentum then implies, on a classical level, that the topological dipole moment is conserved, i.e., $\partial_t D_i = 0$. 

For a texture with a non-zero $Q_{\rm top} \neq 0$, the dipole moment can be invoked to introduce a collective variable  $R_i(t)$ specifying the position of the texture,
\begin{equation}
\label{eq:collectivecoor}
 R_i(t)  = D_i(t)/Q_{\rm top} .
\end{equation}
The conservation of linear momentum can then be expressed in terms of the so-called Thiele equation in the absence of Gilbert damping, $\alpha=0$, \cite{Thiele1973}
\begin{align} \label{eq:Pconservation}
\partial_t \vec P = 4\pi s\, \hat z \times \partial_t \vec D =  \vec G \times \partial_t \vec R = 0,
\end{align}
where the gyrocoupling vector, $\vec G = 4\pi s Q_{\rm top} \hat z$, points in the direction  $\hat z$  normal to the two-dimensional plane.
Of course, if the magnetization couples to additional degrees of freedom, like phonons, and exchanges momentum
with another subsystem, Eq.~\eqref{eq:Pconservation} acquires corrections. In
this case, the time derivative of $\vec P$ is balanced by a
force describing the exchange of linear momentum that
eventually induces also a finite Gilbert damping $\alpha$. The
present work, however, is concerned with the intrinsic
properties of a magnetic system that is decoupled from
the environment.

The notion of linear momentum of a magnetic texture has been discussed for many years \cite{Thiele1976,Haldane1986,Volovik1987}, and it is still intriguing the community \cite{Watanabe2014,Tchernyshov2015,Dasgupta2018,Di2021,seiberg2024}. Indeed, it is a pertinent scientific question whether the Thiele equation \eqref{eq:Pconservation} for the collective coordinate \eqref{eq:collectivecoor} is robust or only holds on the classical level for a rigid texture as thermal or quantum fluctuations might induce corrections. A previous theoretical analysis on the influence of spin wave excitations by Psaroudaki {\it et al.} \cite{Psaroudaki2017} suggests that the equation of motion is renormalized 
\begin{equation}
\vec G\times\frac{\partial \vec  R}{\partial t} = - m_{\rm eff}\frac{\partial^{2}\vec R}{\partial t^{2}} + \cdots ,
\label{eq:withmass}
\end{equation}
and the skyrmion acquires, in particular, a finite mass $m_{\rm eff}$. The ellipses represent further corrections like internal damping and higher-order time derivatives \cite{Schuette2014,Back2020}. Equation \eqref{eq:withmass} is analogous to the dynamics of a classical electron in an external magnetic field $\sim \vec G$ and possesses cyclotron solutions with finite frequency $\omega_c = | \vec G | /m_{\rm eff}$. For the corresponding periodic cyclotron motion, the velocity is finite, $\partial_t \vec R \neq 0$, such that the conservation law \eqref{eq:Pconservation} for linear momentum appears to be violated. 

This seemingly striking conflict between a finite skyrmion mass $m_{\rm eff}$ and the conservation of linear momentum only occurs for the collective coordinate defined in terms of the topological dipole moment \eqref{eq:collectivecoor}. For completeness, we would like to note that other definitions for the collective coordinate $\vec R$ have been also proposed in the literature that are distinct from Eq.~\eqref{eq:collectivecoor} and are not related to the linear momentum of the texture, see \cite{Sheka2006,Moutafis2009,Buttner2015,Wu2022}. In this case, a finite mass $m_{\rm eff}$ quite naturally arises because the collective coordinate $\vec R$ is here not orthogonal to the gapped magnon excitations on the level of linear spin wave theory \cite{Ivanov1989,Makhfudz2012,Kravchuk2018}. 

Being the linear momentum of the magnetic texture, the topological dipole moment and its associated collective coordinate $\vec R$ of Eq.~\eqref{eq:collectivecoor} belong to the low-energy degrees of freedom of the magnetic system. Consequently, the form of the effective equation of motion for $\vec R$ is of fundamental importance for the description of quantum skyrmions as well as quantum phases of many-skyrmion systems. For example, the analysis of a Bose-Einstein condensation of quantum skyrmions carried out in Ref.~\cite{Balents2016} relies on the assumption that the conservation law of Eq.~\eqref{eq:Pconservation} applies, and its conclusions would be invalidated in the presence of a finite $m_{\rm eff}$. The presence of a finite mass $m_{\rm eff}$ substantially complicates the semiclassical quantization of magnetic skyrmions as discussed in detail by Ochoa and Tserkovnyak \cite{Ochoa2019}.

In the present work, we demonstrate that, instead, the conservation law \eqref{eq:Pconservation} remains valid, and the collective coordinate defined in terms of the topological dipole moment rigorously obeys an equation of motion, both on the classical and the quantum level, with 
\begin{equation}
m_{\rm eff}=0,
\label{eq:nomass}
\end{equation}
even if one includes fluctuations. In fact, all the terms on the right-hand side of Eq.~\eqref{eq:withmass} vanish identically for a spin system that does not couple to other degrees of freedom such as lattice vibrations, i.e., internal damping terms are not fluctuation-generated either. We show that this is a consequence of a local dipole conservation law obeyed by the topological charge density 
\begin{equation}
    \partial_{t}\rho_{\rm top}+\partial_{i}\partial_{j}J_{ij}=0,
    \label{eq:dipole_continuity}
\end{equation}
where the current tensor $J_{ij}$ is related to the stress tensor of the problem \cite{Papanicolaou1991}, see also Eq.~\eqref{eq:continuity2} below. 
We establish explicitly that Eq.\eqref{eq:dipole_continuity} holds as an exact equation on the operator level that is free from quantum anomalies, thus generalizing the seminal results of Ref.~\cite{Papanicolaou1991} to the quantum level.
Hence, spin fluctuations that
deform a rigid magnetic texture do not generate inertia.
This conclusion derives from the fact that both
the conservation of the
linear momentum and of the topological dipole moment
continue to be tied to each other, i.e., there is no anomaly
in the quantum field theory that would lead to the violation of the dipole conservation law. 
Finally, we explain how
our results can be reconciled with the previous findings of Psaroudaki {\it et al.}~\cite{Psaroudaki2017} who 
obtained results that seem to
imply a finite value for the mass of a skyrmion.

Despite the vanishing mass, the quantum dynamics of skyrmions is highly nontrivial. In an explicit spin-wave analysis, we illustrate that the vanishing mass is the consequence of the cancellation of the movement of the quasi-classical spin configuration and a cloud of quantum fluctuations.
We further show that the actual quantum dynamics is identical to that of bosonic charges in the lowest Landau level (LL), where the kinetic energy is also quenched. In fact, such an analogy has a deeper root: for the topological charge density of skyrmions, we show that it is subject to the Girvin-MacDonald-Platzman algebra \cite{Girvin1986} that governs neutral intra-LL excitations. Those  are particularly non-trivial for fractional quantum Hall states. Hence, we expect analogies between the problem of quantum skyrmions and fractional quantum Hall physics that may allow for a classification of incompressible skyrmion quantum liquids in analogy to the one achieved for electrons in high magnetic fields.

The dipole conservation law of Eq.~\eqref{eq:dipole_continuity} further allows for a close correspondence between the skyrmion dynamics and the one discussed in fracton theories.
Dipole conservation laws have been discovered in specific quantum spin models featuring both topological orders and quantum spin liquid phases, where they have been associated with emergent $U(1)$ charges of so-called fracton quasiparticles \cite{Chamon2005, Bravyi2011, Haah2011, Yoshida2013, Vijay2015, Vijay2016}. Various variants of fracton theories have been proposed \cite{Pretko1} that lead to a plethora of unusual phenomena, e.g., the restricted mobility of quasiparticles and the associated anomalous hydrodynamics \cite{Morningstar2020, Gromov2020, Zhang2020, Feldmeier2020}, glassiness and non-ergodic behaviors \cite{Pai2019, Khemani2020, Sala2020}, as well as the emergence of tensor gauge theories \cite{Pretko2, Pretko2018gauge}, see Refs.~\cite{Nandkishore2019, PretkoRev, Gromov2024} for reviews.  
Microscopic realizations of these proposed fracton theories and their experimental verifications in materials 
are scarce, such as  defects in elasticity theory \cite{Pretko2018duality,Pretko2019,Gaa2021}, vortices in superfluids \cite{Doshi2021}, or electrons within the lowest LL \cite{Du2022}. Here, we find that magnetic skyrmions provide another realization of fractonic behavior. The versatility and control of magnetic skyrmion materials achieved over the last years turn them into a promising platform for studying the unusual fractonic phenomena, which have been proposed but so far remain largely unexplored experimentally.
 
The article is organized as follows. In Section \ref{sec:TopDipConservationClassical}, we present an extended review of the conservation law of both linear momentum and the topological dipole moment in the framework of the classical continuum field theory. Section \ref{sec:quantum} discusses the generalization of the results for the quantum field theory. In particular, using a path-integral formalism, a Ward-Takahashi identity associated with translation invariance 
is derived, and the continuity equation for the topological density is generalized to the operator level. From this, it follows that the topological dipole moment remains conserved in the quantum field theory. In Section \ref{sec:Spinwavetheory}, we confirm our findings with an explicit calculation using linear spin-wave theory and elucidate the relation to previous works. In Section \ref{sec:Discussion}, we close with a discussion of our findings on (a) the skyrmion mass problem, (b) the correspondence with the lowest Landau level, and (c) the skyrmion-fracton correspondence.

\begin{table}[b]
\caption{Convention for indices in this paper. We use the abbreviation $x = (x_\mu) = (\vec x, t)$ and identify the zeroth component of the spacetime coordinate $x_\mu$ with time, $x_0 = t$.
}
\label{tab:t1}
  \begin{center}
    \begin{tabular}{l l l}
      \hline
      \hline
      Spin vector indices & \hspace{5pt} & $\alpha, \beta, \gamma \in \{ 1, 2, 3\}$\\
      Spatial indices & \hspace{5pt} & $i, j, k, l \in\{ 1, 2\}$ \\
      Spacetime indices & \hspace{5pt} & $\mu, \nu, \lambda \in \{0, 1, 2\}$\\
      \hline
      \hline
    \end{tabular}
  \end{center}
\end{table}

\section{Topological dipole conservation law in classical field theory}
\label{sec:TopDipConservationClassical}
The theory of quantum skyrmions  ultimately requires a quantum theory, which we will develop in Section~\ref{sec:quantum}. Before that, we discuss here the corresponding quasi-classical theory, an approach that will eventually be justified by our finding that the theory is not affected by quantum anomalies. In particular, we will find simple and rather transparent arguments that  reveal the important role of the Girvin-MacDonald-Platzman algebra that was initially developed to   describe  neutral modes of the lowest Landau level  in the fractional quantum Hall problem.
 Following Refs.~\cite{Papanicolaou1991,Du2022,seiberg2024}, we first review the notion of linear momentum for a classical field theory whose dynamics is governed by the Landau-Lifshitz equation. We introduce the field theory as well as the Poisson bracket in Section \ref{subsec:ClassFieldTheory}. In Section \ref{subsec:ConsLinearMomentum}, we discuss the conservation of linear momentum and present two derivations for the continuity equation associated with translational invariance, i.e., from the equation of motion and from Noether's theorem. In particular, we discuss how a manifestly spin-gauge invariant energy-momentum tensor is obtained with the help of an ``improvement transformation". Topological terms and anomalies are shortly addressed in Section \ref{sec:anomalies} before we discuss the continuity equation for the topological charge density $\rho_{\rm top}$ whose form is intimately related to the property of $\rho_{\rm top}$ as a  generator of area-preserving diffeomorphisms. This finally leads to the conservation of the total topological dipole moment of the spin texture. 
Table \ref{tab:t1} summarizes the convention for indices used throughout this work.

\subsection{Classical field theory}
\label{subsec:ClassFieldTheory}

In addition to the topological density of Eq.~\eqref{eq:TopDensity}, the topological current will be playing an important role in the following, 
\begin{align}
\label{eq:topcurrent}
j_{\text{top}, i} &= 
\frac{\epsilon_{ij}}{4\pi} \vn \cdot (\partial_{j} \vn \times \partial_t \vn),
\end{align}
with $\epsilon_{12} = - \epsilon_{21} = 1$. The topological density and its current obey the continuity equation,
\bea
\label{eq:continuity1} 
\partial_t  \rt + \partial_i   j_{\text{top}, i}  = 0.
\eea
This equation has a purely geometric origin due to the constraint $\vec n(\vec x,t)^2 = 1$ of the unit vector field. If this constraint is locally violated  by hedgehog defects in 2+1 spacetime, the right-hand side will not vanish but will be given by the hedgehog density. Throughout this work, we assume, however, that hedgehog defects are energetically suppressed and not present. 

The dynamics of the unit vector field $\vec n(\vec x,t)$ representing the orientation of the magnetization is governed by the Landau-Lifshitz equation of motion
\bea
\label{eq:LLeq}
\partial_t \vn &=& \frac{1}{s} 
\vn \times \frac{\delta \mathcal{W}}{\delta \vn},
\eea
that describes the precession of the magnetic moments. Here, $s = M_s/\gamma$ is a spin density with unit $\hbar$ per volume and can be expressed in terms of the saturation magnetization $M_s$ and the gyromagnetic ratio $\gamma = g\mu_B/\hbar$, where $g$ is the g-factor, $\mu_B$ is the Bohr magneton, and $\hbar$ is the reduced Planck constant. The energy functional
\begin{align}
\mathcal{W} = \int_V d\vec x\, \mathcal{H}(\vec n, \partial_i \vec n)
\end{align}
is assumed to be local with the density $\mathcal{H}$. For this work, the explicit form of $\mathcal{H}$ is not important, but we assume that the theory is invariant with respect to spatial translations, implying that $\mathcal{H}$ does not depend explicitly on the spatial coordinate. For simplicity, we also assume that $\mathcal{H}$ depends only on the field itself and its spatial derivative; however, our results can be straightforwardly generalized to cases where $\mathcal{H}$ also depends on higher-order derivatives. 

The equation of motion \eqref{eq:LLeq} is generated by the Poisson bracket $\partial_t \vn = \{\vn, \mathcal{W} \}$ defined as follows
\begin{align} \label{eq:Poisson}
\{n_{\alpha}(\vec x,t), n_\beta(\vec x',t) \} = -\frac{1}{s} \epsilon_{\alpha\beta \gamma} n_{\gamma}(\vec x,t) \delta(\vec x - \vec x').   
\end{align}
A central role is played by the Poisson bracket of the topological charge density and the field $\vec n$
\begin{align} \label{eq:PoissonTopChargeField}
&\{\rt(\vec x,t), \vec n(\vec x',t) \} = 
\frac{1}{4\pi s} \epsilon_{ij} 
\partial_j \delta(\vec x-\vec x') \partial_i \vec n(\vec x,t),
\end{align}
which identifies $\rt$ as a generator of area-preserving diffeomorphisms \cite{Du2022}, as we will further explain in Section \ref{subsec:conservation charge}. For the Poisson bracket of topological charge densities at different spatial positions, we then immediately obtain 
\begin{align} \label{eq:PoissonTopCharge}
&\{\rt(\vec x,t), \rt(\vec x',t) \} = 
\frac{1}{4\pi s} \epsilon_{ij} 
\partial_j \delta(\vec x-\vec x') \partial_i \rt(\vec x,t).
\end{align}
Remarkably, this corresponds to the Girvin-MacDonald-Platzman algebra in the long-wavelength limit \cite{Girvin1986}, which suggests a close relationship with the physics of the lowest Landau level \cite{Papanicolaou1991,Du2022}. From Eq.~\eqref{eq:PoissonTopCharge} follows the Poisson bracket of the topological dipole moments \eqref{eq:TopDipole}
\begin{align} \label{eq:PoissonTopDip}
\{D_i, D_j\} =  \frac{Q_{\rm top}}{4\pi s} \epsilon_{ji} .
\end{align}
Their Poisson bracket does not vanish in the presence of a topological texture with a nonzero $Q_{\rm top}$.

The action $S = \int dt \int_V d\vec x \mathcal{L}$ of the classical field theory is defined by the Lagrangian density  \cite{auerbachBook}
\bea \label{eq:Lagrangedensity}
\mathcal{L} = s \vec A(\vec n) \partial_t \vec n - \mathcal{H},
\eea
where the first term represents the Berry phase with the vector potential satisfying $\frac{\partial}{\partial \vec n} \times \vec A = \vec n$. The choice of $\vec A$ is not unique; the equations of motions are invariant with respect to the spin-gauge transformation $\vec A \to \vec A + \frac{\partial}{\partial \vec n} \chi(\vec n)$ with some smooth function $\chi(\vec n)$. 
Its Euler-Lagrange equations reproduce the Landau-Lifshitz equations. When evaluating them, it is important to impose the local constraint $\vec n^2 = 1$, e.g.~with the help of a Lagrange multiplier. This constraint requires that the variation of the action is projected onto the subspace perpendicular to $\vec n$. The effective Euler-Lagrange equations thus read
\begin{align} \label{eq:SaddlePoint}
(\mathbb{1} - \vec n \vec n^T) \frac{\delta S}{\delta \vec n} = (\mathbb{1} - \vec n \vec n^T) \Big(\frac{\partial \mathcal{L}}{\partial \vec n} - \partial_\mu \frac{\partial \mathcal{L}}{\partial \partial_\mu \vec n}\Big) = 0,
\end{align}
that is equivalent to Eq.~\eqref{eq:LLeq}.

\subsection{Conservation law for linear momentum}
\label{subsec:ConsLinearMomentum} 

The notion of linear momentum for a field $\vec{n}$, that precesses according to the Landau-Lifshitz equation, has been extensively discussed in the literature \cite{Thiele1976,Haldane1986,Volovik1987,Papanicolaou1991,Tchernyshov2015,Dasgupta2018,Di2021,seiberg2024}. The linear momentum following from Noether's theorem is not manifestly spin-gauge independent, which prompted many of the discussions. 
Circumventing this problem, Papanicolao and Tomaras \cite{Papanicolaou1991} proposed a solution by starting instead from the Landau-Lifshitz equation of motion without any reference to the Lagrangian density of Eq.~\eqref{eq:Lagrangedensity}. 
This derivation is reviewed in Section \ref{sec:DerivationFromLandauLifshitz} for the convenience of the reader. In Section \ref{sec:Noether}, the same result is rederived using Noether's theorem. In particular, we demonstrate explicitly how the spin-gauge-dependent energy-momentum tensor can be ``improved" by a transformation such that the result of Papanicolao and Tomaras for the linear momentum is recovered.

\subsubsection{Derivation from the Landau-Lifshitz equation}
\label{sec:DerivationFromLandauLifshitz}

Multiplying the Landau-Lifshitz equation \eqref{eq:LLeq} first with $\vec n \times$ and afterwards with $\partial_i \vec n$, i.e., the generator of translations $\partial_i$ applied to the field $\vec n$, we obtain  
\bea \label{eq:LLGderivation1}
\partial_i \vec n\cdot(\vec n \times \partial_t \vec n) = -\frac{1}{s} \partial_i \vec n \cdot \frac{\delta \mathcal{W}}{\delta \vec n},
\eea
where we have used the fact that the unit vector field obeys $(\partial_i \vec n)\vec n = 0$. Due to translational invariance, the left-hand side can be expressed in terms of a divergence 
$\partial_i \vec n \frac{\delta \mathcal{W}}{\delta \vec n} =  \partial_j \sigma_{ji}$, where 
\bea \label{eq:stressT}
\sigma_{ji} =  \delta_{ji} \mathcal{H} - \frac{\partial \mathcal{H}}{\partial \partial_j \vn} \cdot \partial_i \vn 
\eea
is the stress tensor associated with the static part of the action. Identifying the right-hand side of Eq.~\eqref{eq:LLGderivation1} with the topological current \eqref{eq:topcurrent}, we arrive at 
\begin{align}
\label{eq:momentumConservation}
4\pi s\, \epsilon_{ki} j_{\text{top}, k} &= \partial_j \sigma_{ji}.
\end{align}
This equation is intimately related to the translational invariance of the theory and is associated with the conservation of linear momentum. In order to bring it into the standard form of a continuity equation, we exploit the identity
\begin{align} \label{eq:IdTopCurrent}
j_{\text{top}, i} &= \partial_j (x_i j_{\text{top}, j}) - x_i \partial_j j_{\text{top}, j} \\
&= \partial_j (x_i j_{\text{top}, j}) + x_i \partial_t \rt,
\end{align}
where we used Eq.~\eqref{eq:continuity1} in the second line. Plugging this into Eq.~\eqref{eq:momentumConservation}, we obtain a continuity equation for the linear momentum,
\begin{align} \label{eq:LinearMomentumContinuityEq}
\partial_t p_i + \partial_j \Pi_{ji} = 0,
\end{align}
where the linear momentum density and its current, respectively, are given by 
\begin{align} \label{eq:Momentum}
p_i &=  4\pi s\,  \epsilon_{ki} x_k \rt,\\
\label{eq:MomentumCurrent}
\Pi_{ji} &=  - \sigma_{ji} 
+ 4\pi s\, \epsilon_{ki} x_k j_{\text{top}, j}.
\end{align}
The total linear momentum thus defined, 
\begin{align} \label{eq:TotMomentum}
P_i  = \int d \vec x\, p_i = 4\pi s\,  \epsilon_{ki} D_k, 
\end{align}
is proportional to the topological dipole moment, and it is indeed a generator of translations in the sense that its Poisson bracket with the field $\vec n$ is given by
\begin{align}
  \partial_{i} \vec n = -\{\vec n,P_i\}.
\end{align}
In the next section, we will revisit this derivation from the perspective of Noether's theorem.

\subsubsection{Derivation from Noether's theorem}
\label{sec:Noether}

In order to obtain the conservation law for linear momentum using Noether's theorem, we consider a local infinitesimal translation in 2+1 space time by the amount $\xi_\mu(x)$ that  leads to a change of the field configuration 
\bea \label{eq:Transformation}
\delta \vec n(x) = - \xi_\mu(x) \partial_\mu \vec n(x).
\eea
This change in turn results in a modification of the action 
\begin{align} \label{eq:variationS2}
\delta S &= -\int dx \  \left(\frac{\partial \mathcal{L}}{\partial \vec n}  \xi_\mu \partial_\mu \vec n + \frac{\partial \mathcal{L}}{\partial \partial_\nu \vec n} \partial_\nu ( \xi_\mu \partial_\mu \vec n) \right)
\\ &=-
\int dx \ \left( \xi_\mu \frac{\delta S}{\delta \vec n} \partial_\mu \vec n + \partial_\nu \left(  \frac{\partial \mathcal{L}}{\partial \partial_\nu \vec n}\xi_\mu \partial_\mu \vec n \right)\right).
\end{align}
The second term is only a surface term. The vanishing of $\delta S$ for all choices of $\xi_\mu(x)$ thus requires that the first term vanishes locally. With the identification for a translationally invariant theory $\frac{\delta S}{\delta \vec n} \partial_\mu \vec n = - \partial_\nu T_{\nu \mu}$, this amounts to the conservation law
\bea \label{eq:FullContinuity}
\partial_{\mu} T_{\mu\nu} = 0,
\eea
with energy-momentum tensor
\bea
\label{eq:emtensor}
T_{\mu\nu} &=& \frac{\partial \cl}{\partial \partial_{\mu}\vn} \cdot \partial_{\nu}\vn - \delta_{\mu\nu} \cl.
\eea

The time component of Eq.~\eqref{eq:FullContinuity} represents the continuity equation for the energy density $T_{00} = \mathcal{H}$ and its associated energy current $T_{i0}$. The spatial component  describes the conservation of linear momentum, $\partial_\mu T_{\mu i} = 0$, where in particular
\begin{align}
T_{0i} &= s \vec A \cdot \partial_i \vec n, \qquad
T_{ji} = \sigma_{ji} - \delta_{ji} s \vec A \cdot \partial_t \vec n.
\end{align}
The fact that $T_{0i}$ depends on the spin-vector potential $\vec A$ and is thus not manifestly spin-gauge invariant has caused many discussions \cite{Thiele1976,Haldane1986,Volovik1987,Papanicolaou1991,Tchernyshov2015,Dasgupta2018,Di2021}. However, it is important to notice that the spatial current $T_{ji}$ also depends on $\vec A$.
Although $T_{\mu\nu}$ is not spin-gauge invariant, the conservation law is, in fact, manifestly invariant. Using 
\bea 
\partial_t (\vec A \cdot\partial_i \vec n) - \partial_i (\vec A \cdot\partial_t \vec n) = \vec n \cdot (\partial_t \vec n \times \partial_i \vec n),
\eea
we obtain 
\bea \label{eq:NoetherMomentumConservation}
\partial_t T_{0i} + \partial_j T_{ji} =  4\pi s\, \epsilon_{i j} j_{\text{top}, j} + \partial_j \sigma_{ji} = 0,
\eea
that is equivalent to Eq.~\eqref{eq:momentumConservation}. The ``improved" spin-gauge invariant momentum \eqref{eq:Momentum} and its current \eqref{eq:MomentumCurrent} are formally related to $T_{\mu i}$ by the transformation \cite{seiberg2024}
\begin{align}
p_i &= - T_{0i} + \partial_j U_{ji} , \\
\Pi_{ji} &= - T_{ji} - \partial_t U_{ji} + \epsilon_{kj} \partial_k V_{i} ,
\end{align}
where
\begin{align}
U_{ji} &= 
s \epsilon_{kj} \epsilon_{il} x_l \vec A \partial_k \vec n ,
\quad
V_i = s \epsilon_{il} x_l \vec A \partial_t \vec n .
\end{align}

\subsection{Topological terms and anomalies}
\label{sec:anomalies}

Recently, the modifications of the conservation law for linear momentum due to topological terms and anomalies have been addressed by Seiberg \cite{seiberg2024}. An additional term in the action of the form
\begin{align}
\mathcal{L}_{\rm top} = - 4\pi s\, \epsilon_{ki} x^0_k j_{\text{top},i},
\end{align}
with some constants $x^0_k$, does not modify the equations of motion, and in this sense it is topological. It leads to an additional contribution to the energy-momentum tensor
\begin{align}
T^{\rm top}_{\mu i} = 4\pi s\, \epsilon_{ki} x^0_k j_{\text{top},\mu},
\end{align}
where the time component $j_{\text{top},0} = \rt$. This additional contribution does not modify the conservation law because $\partial_\mu T^{\rm top}_{\mu i} = 4\pi s\,\epsilon_{ki} x^0_k \partial_\mu j_{\text{top},\mu} = 0$ vanishes due to Eq.~\eqref{eq:continuity1}. Formally, it modifies however the momentum density and its current, and we get instead of Eqs.~\eqref{eq:Momentum} and \eqref{eq:MomentumCurrent},
\begin{align} \label{eq:Momentum2}
p_i &=  4\pi s\,  \epsilon_{ki} (x_k - x^0_k) \rt,\\
\label{eq:MomentumCurrent2}
\Pi_{ji} &=   - \sigma_{ji} 
+ 4\pi s\, \epsilon_{ki} (x_k - x^0_k) j_{\text{top}, j}.
\end{align}
The constants $x_i^0$ thus specify the choice of the origin with respect to which the topological dipole density and the associated current are measured. The topological dipole moment $D_i$ and the total linear momentum are thus not uniquely defined for a non-zero $Q_{\rm top}$ but depend on the choice of the coordinate system.

Moreover, it was pointed out that linear momentum is not conserved for a finite system with periodic boundary conditions \cite{seiberg2024}. Defining the spatial coordinate as $x_i \in [0,L_i)$ for a finite system with linear size $L_i$ along the $i$-axis, we get 
$\partial_j x_i = \delta_{ij} (1 - L_j \delta(x_j))$, where the Einstein summation convention is not assumed here. In the rest of this Section \ref{sec:anomalies}, we will not assume the Einstein convention, but we return to the convention elsewhere.
Equation~\eqref{eq:IdTopCurrent} is then modified into
\begin{align}
j_{\text{top}, i} &= \sum_j \partial_j (x_i j_{\text{top}, j}) + L_i \delta(x_i) j_{\text{top}, i} + x_i \partial_t \rt.
\end{align}
Consequently, the continuity equation acquires a source term,\footnote{See Eqs.~(46), (47) and (107) of Ref.~\cite{seiberg2024}} 
%
\begin{align}
\partial_t p_i + \sum_j \partial_j \Pi_{ji} = 4\pi s\, \sum_j \epsilon_{ij} \delta(x_j) L_j  j_{\text{top}, j},
\label{eq:continuityPBC}
\end{align}
that is only finite at the boundary of the system, $x_j = 0$. 
In the present work, we are rather interested in the bulk of the material with open boundary condition where such boundary terms drop out of Eq.~\eqref{eq:continuityPBC}, and we get back Eq.~\eqref{eq:LinearMomentumContinuityEq}.

\subsection{Conservation law for the topological charge density}
 \label{subsec:conservation charge}

Due to the form of its Poisson bracket \eqref{eq:PoissonTopChargeField}, the topological charge $\rt$ is a generator of area-preserving diffeomorphisms \cite{Du2022}, i.e., local spatial transformations that are area-preserving, $\partial_i \xi_i = 0$,
\begin{align} \label{eq:generatorAPD}
4\pi s \int d\vec x' \lambda(\vec x',t)  &\{  \rt(\vec x',t), \vec n(\vec x,t) \} = 
\\\nonumber &\quad - (\epsilon_{ij} \partial_j \lambda(\vec x,t)) \partial_i\vec n(\vec x,t),
\end{align}
cf.~Eq.~\eqref{eq:Transformation} with $\xi_i(\vec x, t) = \epsilon_{ij} \partial_j \lambda(\vec x,t)$. Similarly, applying this generator to the energy functional $\mathcal{W}$ gives 
\begin{align}
4\pi s \int d\vec x' \lambda(\vec x',t)  &\{  \rt(\vec x',t), \mathcal{W} \} = 
\\\nonumber &\quad 
- \int d \vec x (\epsilon_{ki} \partial_i \lambda(\vec x,t)) \partial_j \sigma_{jk}.
\end{align}
Variation with respect to $\lambda$ and identifying $\partial_t \rt = \{  \rt, \mathcal{W} \}$ yields the conservation law for the topological charge density
\bea
\label{eq:continuity2}
\partial_t \rt + \frac{1}{4\pi s} \epsilon_{ik} \partial_i \partial_j \sigma_{jk} &=& 0.
\eea
It has the general form of the dipole conservation law of  Eq.~\eqref{eq:dipole_continuity} in the introduction where the tensor current  $J_{ij}=\frac{1}{4\pi s} \epsilon_{ik}  \sigma_{jk}$ is related to the stress tensor.
This equation can also be  obtained by combining Eq.~\eqref{eq:momentumConservation} with the continuity equation for the topological current Eq.~\eqref{eq:continuity1}. Note however that Eq.~\eqref{eq:continuity2} is not equivalent to the conservation law of linear momentum Eq.~\eqref{eq:LinearMomentumContinuityEq} because only the 
curl of Eq.~\eqref{eq:momentumConservation} is required to arrive at Eq.~\eqref{eq:continuity2}, reflecting the restriction to area-preserving transformations $\xi_i(x) = \epsilon_{ij} \partial_j \lambda(x)$.

Equation \eqref{eq:continuity2}
and its implications were first derived and examined in Ref.~\cite{Papanicolaou1991}. The double-divergence form allows one to express the time derivative of the topological charge $Q_{\rm top}$ as well as that of the first moment of the topological charge \eqref{eq:TopDipole} as surface integrals, for example,
\begin{align}
\partial_t Q_{\rm top} &=- \frac{1}{4\pi s} \oint_{\partial V} \!\!\!  \td S \hat{N}_j \epsilon_{ik} \partial_i \sigma_{jk}, 
\\
\partial_t D_i &=\frac{1}{4\pi s} \oint_{\partial V} \!\!\! \td S \hat{N}_j \left(-x_i \epsilon_{lk}\partial_l \sigma_{jk} + \epsilon_{jk} \sigma_{ik} \right),
\end{align}
where $\hat{N}$ is the unit vector normal to the boundary surface $\partial V$. With an appropriate boundary condition where the surface integrals vanish, the topological charge $Q_{\rm top}$ and its first moment $D_i$ become conserved quantities. The latter can be identified with the total linear momentum of the system, see Eq.~\eqref{eq:TotMomentum}.

With the help of Eq.~\eqref{eq:continuity2}, one finds that the time derivative of the second moment $\int_V d\vec x\, \vec x^2 \rt$ can be also expressed in terms of a surface integral provided that the stress tensor $\sigma_{ji}$ is symmetric. In this case, the theory is invariant with respect to spatial rotations, and the second moment is related to the net orbital angular momentum as discussed in Ref.~\cite{Papanicolaou1991}. In the following, we focus however mostly on the first moment, i.e. the topological dipole. 

\section{Topological dipole conservation law in quantum field theory}
\label{sec:quantum}

In the quantum field theory, conservation laws due to symmetries in general manifest themselves in operator-valued continuity equations. In this section, we formulate a micromagnetic quantum theory using a path-integral approach. We continue to assume no hedgehog defects in 2+1 spacetime and derive a Ward-Takahashi identity associated with translational invariance  with a focus on area-preserving diffeomorphisms. We show that there are no anomalies, so the conservation of the topological charge and its first moment also hold at the quantum level. 

\subsection{Quantum field theory}
We define the quantum field theory by the partition function expressed in terms of the path integral
\bea
\label{eq:partitionFunc}
\cz &=& \int \cd \vn\ \delta(\vn^2 - 1) \exp\left(-S_E[\vn]\right),
\eea
where the delta function imposes the unit-norm constraint for $\vn$, and the Euclidean action is given by $S_E = \int_0^\beta d\tau \int_V d \vec x \mathcal{L}_E$ where $\mathcal{L}_E = -\frac{s}{\hbar} \vec A i \partial_\tau \vec n + \mathcal{H}$ with imaginary time $\tau = i t/\hbar$ in units of inverse energy and $\partial_\tau = -i\hbar \partial_t$, where $\hbar$ is the reduced Planck constant. The path integral is restricted to paths that are periodic in imaginary time, i.e., with boundary conditions $\vec n(\vec x, \tau)|_{\tau = 0} = \vec n(\vec x, \beta)$.
As usual, an expectation value of an observable $\mathcal{A}(x)$, here with $x = (\vec x, \tau)$, is defined as
\begin{align}
\langle \mathcal{A}(x) \rangle = \!\frac{1}{\cz}\!
\int \cd \vn \delta(\vn^2 -1) \exp(-S_E[\vn]) \mathcal{A}(x)  .
\end{align}
This coincides with the expectation value in the operator formalism $\langle \mathcal{A}(x) \rangle = \langle \hat{\mathcal{A}}(x) \rangle$ for an operator $\hat{\mathcal{A}}$. Evaluating an expectation value of two observables at distinct points in spacetime, one obtains a correlation function $\langle  \mathcal{A}(x) \mathcal{B}(x') \rangle$.  This object is related to the time-ordered correlation function in the operator formalism by 
$\langle  \mathcal{A}(x) \mathcal{B}(x') \rangle = \langle T \{ \hat{\mathcal{A}}(x) \hat{\mathcal{B}}(x') \} \rangle$, 
where $T$ is the time-ordering operator. For later reference, we note that the time derivative for bosonic fields $\mathcal{A}$ and $\mathcal{B}$ obeys
\begin{align} \label{eq:TimeDerCorr}
&\partial_\tau \langle T \{ \hat{\mathcal{A}}(\vec x, \tau) \hat{\mathcal{B}}(\vec x',\tau') \} \rangle =
\\\nonumber & 
\,\langle T \{ \partial_\tau \hat{\mathcal{A}}(\vec x, \tau) \hat{\mathcal{B}}(\vec x',\tau') \} \rangle 
+\delta(\tau-\tau') 
\langle [\hat{\mathcal{A}}(\vec x, \tau) , \hat{\mathcal{B}}(\vec x',\tau) ] \rangle
\end{align}
with the commutator $[\hat{\mathcal{A}}, \hat{\mathcal{B}}] = \hat{\mathcal{A}} \hat{\mathcal{B}} - \hat{\mathcal{B}}\hat{\mathcal{A}}$.\\

\subsection{Ward-Takahashi identity for the topological charge density}
\label{sec:subSD}

The aim of this section is the derivation of a Ward-Takahashi identity for the topological density $\rt$. 
The derivation follows standard methods \cite{Peskin}.
We start by considering the generating functional
\begin{align}
\label{eq:eqZ0}
\cz[\vec h] &=& \int \cd \vn \ \delta(\vn^2 -1) \exp(-S_E[\vn] + \int dx\, \vec h \vec n),
\end{align}
that is a functional of the field $\vec h$, and we introduced the abbreviation $\int dx = \int_0^\beta d\tau \int_V d \vec x$.
Following Section \ref{sec:Noether}, we consider the change of the field configuration: $\vec n(x) \to \vec n(x) + \delta \vec n(x)$, where
\begin{align}
\delta \vec n(x) = - \xi_i(x) \partial_i \vec n(x).
\end{align}
\begin{widetext}
This transformation induces a change $\cz \to \cz + \delta\cz$, where we get up to first order in $\xi_i$
\begin{align} \label{eq:deltaZ}
\delta\cz[\vec h] = \int \cd \vn\,  \delta(\vn^2 -1) \exp(-S_E[\vn]+ \int dx\, \vec h \vec n) 
\int \td x\, \left[ \xi_i \partial_i \vn \frac{\delta S_E}{\delta \vn} + \partial_{\mu} \left(\xi_i \partial_i \vn \frac{\partial \cl_E}{\partial \partial_{\mu}\vn}\right) -  \xi_i \vec h(x) \partial_i \vec n(x) \right] .
\end{align}
The Jacobian of the transformation is unity  to linear order in $\xi_i$, which is a standard result for the consideration of Ward-Takahashi identities associated with translation symmetries\footnote{In more detail, the Jacobian for a change of variables is formally given by the determinant of an infinite-dimensional matrix, $|\det[\delta(x - x') - \xi_i(x) \partial_i \delta(x-x')]|$, where the matrix indices, $x$ and $x'$, correspond to spacetime coordinates. Using the identity $\det(I + \epsilon B) = 1 + \epsilon \tr(B) + \co(\epsilon^2)$ and the fact that the derivative $\partial_i$ has vanishing diagonal matrix elements within a lattice-regularized scheme, one can show that the Jacobian of the transformation is indeed unity at linear order, so the measure of the path integral does not change.}. 
Moreover, as the change $\delta \vn$ is orthogonal to the field, $\delta \vn\cdot \vn =0$, the argument of the delta function in $\cz$ does not change either at this order. 
The terms in the square bracket thus derive only from the change of the exponent.  

The final step is to consider a specific form for the function $\xi_i(x)$. We require that it vanishes at the spacetime boundary such that the second term in the brackets of Eq.~\eqref{eq:deltaZ} does not contribute. In addition, we consider an area-preserving transformation $\xi_i(x) = \epsilon_{il} \partial_l \lambda(x)$. Performing an integration by parts, we then obtain
\begin{align}
\delta\cz[\vec h] =  \int \cd \vn\,  \delta(\vn^2 -1) \exp(-S_E[\vn]+ \int dx\, \vec h \vec n) 
\int \td x\, \lambda(x) \left[ - \epsilon_{il} \partial_l \partial_i \vn \frac{\delta S_E}{\delta \vn}  +  \epsilon_{il} \partial_l (\vec h(x) \partial_i \vec n(x)) \right] .
\end{align}
Following the arguments of Section \ref{sec:TopDipConservationClassical}, we can identify
\begin{align}
-\epsilon_{il} \partial_l \Big(\partial_i \vec n(x) \frac{\delta S_E}{\partial \vec n(x)}\Big) = \frac{4\pi s}{\hbar} i \partial_\tau \rt(x) + \epsilon_{ik} \partial_i \partial_j \sigma_{jk}(x).
\end{align}
As the transformation should leave $\cz[\vec h]$ invariant, i.e., $\delta\cz[\vec h] = 0$ for any $\lambda(x)$, we finally obtain the Ward-Takahashi identity
\bea \label{eq:SD}
\Big\langle \exp(\int dx'\, \vec h(x') \vec n(x')) \Big[
 \frac{4\pi s}{\hbar} i \partial_\tau \rt(x) + \epsilon_{ik} \partial_i \partial_j \sigma_{jk}(x) + \epsilon_{il} \partial_l (\vec h(x) \partial_i \vec n(x)) \Big]
 \Big\rangle = 0,
\eea
that is the central result of this section. Putting the field $\vec h$ to zero, we arrive at the expectation value of the conservation law \eqref{eq:continuity2} in the operator formalism,
\begin{align}
\langle  \frac{4\pi s}{\hbar} i \partial_\tau \hat{\rho}_{\text{top}}(x) + \epsilon_{ik} \partial_i \partial_j \hat \sigma_{jk}(x) \rangle = 0.
\end{align}
Taking the functional derivative $\frac{\delta}{\delta h_\alpha(x')}$ of Eq.~\eqref{eq:SD}, putting $\vec h = 0$ afterwards, and comparing with the general relation \eqref{eq:TimeDerCorr}, we can identify the time-ordered correlation function in the operator formalism
\bea 
\langle T \{ \Big(\frac{4\pi s}{\hbar} i \partial_\tau \hat{\rho}_{\text{top}}(x) + \epsilon_{ik} \partial_i \partial_j \hat \sigma_{jk}(x)  \Big) \hat n_{\alpha}(x') \} \rangle = 0,
\eea
provided that the commutator obeys
\bea
[\hat{\rho}_{{\rm top}}(\vec x, \tau), \hat{n}_\alpha (\vec x',\tau)] = i\hbar  \frac{1}{4\pi s} 
\epsilon_{il} \partial_l \delta(\vec x-\vec x') \partial_i \hat{n}_\alpha(\vec x, \tau)).
\eea
Compared with the classical Poisson bracket \eqref{eq:PoissonTopChargeField}, this commutator indeed has the expected form with the identification $\{\mathcal{A},\mathcal{B}\} \to \frac{1}{i\hbar} [\hat{\mathcal{A}},\hat{\mathcal{B}}]$ when going from classical fields to quantum operators. 
Taking further functional derivatives of Eq.~\eqref{eq:SD} and using the general rules of calculation involving Poisson brackets and commutators, we can show that 
\bea 
\langle T \{ \Big(\frac{4\pi s}{\hbar} i \partial_\tau \hat{\rho}_{\text{top}}(x) + \epsilon_{ik} \partial_i \partial_j \hat \sigma_{jk}(x)  \Big) \hat n_{\alpha_1}(x'_1) \hat n_{\alpha_2}(x'_2) ... \hat n_{\alpha_n}(x'_n) \} \rangle = 0
\eea
also holds for $n$ field insertions, which essentially implies that the conservation law also holds on the operator level  
\bea \label{eq:ConservationLawOperatorLevel}
\frac{4\pi s}{\hbar} i \partial_\tau \hat{\rho}_{\text{top}}(x) + \epsilon_{ik} \partial_i \partial_j \hat \sigma_{jk}(x) = 0.
\eea
We also find that Eq.~\eqref{eq:PoissonTopCharge} generalizes to the quantum level,
\bea \label{eq:GirvinMacDonaldPlatzman}
[\hat{\rho}_{{\rm top}}(\vec x, \tau), \hat{\rho}_{{\rm top}}(\vec x',\tau)] = i\hbar  \frac{1}{4\pi s} \epsilon_{ij} \partial_i \hat{\rho}_{{\rm top}}(\vec x, \tau) \partial_j \delta(\vec x-\vec x'),  
\eea
which corresponds to 
the Girvin-MacDonald-Platzman operator algebra \cite{Girvin1986} as already alluded to above.
\end{widetext}

\subsection{Conservation of the topological dipole moment}
\label{sec:consSD}

We now discuss the implications for the conservation of topological charge and its dipole moment. 
Performing a spatial integration $\int_V d\vec x ...$ on Eq.~\eqref{eq:ConservationLawOperatorLevel}, we obtain
\begin{align}
\frac{4\pi s}{\hbar} i \partial_\tau \hat{Q}_{\text{top}}(\tau) + \oint_{\partial V} dS
\hat N_j \epsilon_{ik} \partial_i \hat \sigma_{jk}(x) = 0.
\end{align}
If we restrict ourselves to a Hilbert space with states on which the surface operator vanishes, we obtain the conservation of the total topological charge $\partial_{\tau} \hat{Q}_{\rm top}(\tau) =0$.
The conservation of the topological dipole moment can be similarly obtained 
\begin{align} \label{eq:TopDipQuantum}
&\frac{4\pi s}{\hbar} i \partial_\tau \hat{D}_i(\tau) + \\\nonumber
&\quad \oint_{\partial V} dS \hat{N}_m \Big( x_i  \epsilon_{lk} \partial_l  \hat \sigma_{mk}(x) - \epsilon_{mk} \hat \sigma_{ik}(x) \Big)
  = 0.
\end{align}
If the surface operators vanish, we obtain that the topological dipole moment is conserved on the operator level, $\partial_{\tau} \hat{D}_i(\tau) =0$. Integration of Eq.~\eqref{eq:GirvinMacDonaldPlatzman} also yields the generalization of Eq.~\eqref{eq:PoissonTopDip} to the quantum level,
\bea
\label{eq:Dcommutator}
\left[\hat{D}_i, \hat{D}_j\right] &=& i\hbar \frac{\hat Q_{\rm top} }{4\pi s} \epsilon_{ji}    .
\eea
We also get $[\hat{D}_i,  \hat Q_{\rm top}] = 0$.

\section{Topological dipole conservation law in linear spin-wave theory}
\label{sec:Spinwavetheory}

In the last section, we generalized the conservation law for the topological density \eqref{eq:continuity2} to the quantum level, from which it follows that the total topological dipole moment is conserved for appropriate boundary conditions. This implies that Eq.~\eqref{eq:Pconservation} is robust with respect to fluctuations and that, in particular, a skyrmion does not acquire a mass, i.e. $m_{\rm eff} = 0$. It is instructive to confirm this result with an explicit calculation taking into account fluctuations around a spin texture in lowest order using linear spin-wave theory.

For this purpose, we consider a static topologically non-trivial classical texture $\vec n_{\rm cl}(\vec x-\vec R_{\rm cl})$ with constant $\vec R_{\rm cl}$ that fulfils the Landau-Lifshitz equation \eqref{eq:LLeq}. The existence of such a solution requires that the energy functional $\mathcal{W}$ obeys certain properties, which we assume in the following. 
Due to translation invariance, the energy of the texture is independent of $\vec R_{\rm cl}$, which can be identified with a 
classical collective coordinate specifying the position of the classical texture. We assume that the texture at spatial infinity is oriented along a common direction, e.g., along the $z$-axis, $\vec n_{\rm cl}(\vec x) \to \hat z$ for $|\vec x| \to \infty$, and that it possesses a non-zero topological charge $Q_{\text{top},\rm cl} = \int d\vec x\, \rho_{\text{top},\rm cl} \neq 0$ with the classical topological charge density $\rho_{\text{top}, \rm cl} = \frac{1}{4\pi} \vec n_{\rm cl} (\partial_x \vec n_{\rm cl} \times \partial_y \vec n_{\rm cl})$. The skyrmion texture of Fig.~\ref{fig:skyrmion} is an example with $Q_{\text{top},\rm cl} = -1$. 

We assume that the constant $\vec R_{\rm cl}$ has been chosen such that the intrinsic topological dipole moment vanishes, $\int d\vec x\, (x_i - R_{{\rm cl},i}) \rho_{\text{top},\rm cl} = 0$, which can be always achieved as $Q_{\text{top},\rm cl} \neq 0$. In this case, the classical topological dipole moment,
\begin{align} \label{eq:ClassDipMom}
D_{{\rm cl},i} &= \int d\vec x\, x_i\, \rho_{\text{top},\rm cl} = 
R_{{\rm cl},i} Q_{\text{top},\rm cl} ,
\end{align}
is proportional to $\vec R_{\rm cl}$. At this classical level, the dipole moment of the texture is also independent of time, $\partial_t \vec D_{{\rm cl}} = 0$. In the following, we discuss the modifications that arise in the presence of fluctuations around the classical solution using linear spin-wave theory.

\subsection{Holstein-Primakoff representation in the presence of a texture}

For the description of the spin-wave fluctuations, we use the standard Holstein-Primakoff representation of the unit vector field $\vec n(\vec x,t)$. For this purpose, we introduce the static but spatially dependent orthonormal dreibein, $\hat e_1 \times \hat e_2 = \hat e_3$ with $\hat e_i \hat e_j = \delta_{ij}$, where $\hat e_3(\vec x - \vec R_{\rm cl} ) = \vec n_{\rm cl}(\vec x-\vec R_{\rm cl})$ coincides with the classical texture. It is also convenient to introduce the chiral vectors $\hat e_\pm = \frac{1}{\sqrt{2}} (\hat e_1 \pm i \hat e_2)$.
The Holstein-Primakoff representation reads
\begin{align} \label{eq:HolsteinPrimakoff}
\vec n(\vec x,t) &= \hat e_3 \Big(1 - \frac{\hbar}{s}\psi^* \psi \Big) 
\\\nonumber & \quad +
\sqrt{1 - \frac{1}{2}\frac{\hbar}{s} \psi^* \psi} \sqrt{\frac{\hbar}{s}} 
(\hat e_-  \psi^* + \hat e_+ \psi ),
\end{align}
where $\hat e_i = \hat e_i(\vec x- \vec R_{\rm cl})$, and $\psi = \psi(\vec x- \vec R_{\rm cl},t)$ is a complex-valued time-dependent field, and $\psi^*$ is its complex conjugate. Importantly, the choice of $\hat e_1$ and $\hat e_2$ is not unique as they are only defined up to a local rotation around $\hat e_3$. Such a local rotation by an angle $\phi(\vec x, t)$ can be absorbed by a $U(1)$ gauge transformation on the wave function $\psi$ such that Eq.~\eqref{eq:HolsteinPrimakoff} remains invariant,
\begin{align} \label{eq:U(1)}
\hat e_\pm \to \hat e_\pm e^{\mp i \phi},\quad  \psi \to \psi e^{i\phi},\quad  \psi^* \to \psi^* e^{-i\phi} .
\end{align}
Under such a transformation, the spin connection $\hat e_1 \partial_i \hat e_2$ transforms like a vector potential,
\begin{align} 
\hat e_1 \partial_i \hat e_2 = -i \hat e_- \partial_i \hat e_+  \to \hat e_1 \partial_i \hat e_2 - \partial_i \phi .
\end{align}
Its curl is related to the topological charge density via the Mermin-Ho relation \cite{MerminHo1976,Kamien2002}
\begin{align} \label{eq:MerminHo}
\frac{1}{4\pi}\epsilon_{ji} \partial_j (\hat e_1 \partial_i \hat e_2) = \rho_{\text{top},{\rm cl}}  + \rho_{\rm sing}. 
\end{align}
In general, it consists of a smooth part given by $\rho_{\text{top},{\rm cl}}$
as well as of a singular part $\rho_{\rm sing}$,
that arises from the non-differentiability of the unit vectors, $\frac{1}{4\pi}\epsilon_{ji} (\hat e_1 \partial_j \partial_i \hat e_2) = \rho_{\rm sing}$. In principle, the singular part $\rho_{\rm sing}$ can be absorbed by singular gauge transformations.
However, if we choose a dreibein that becomes constant at spatial infinity, e.g., $\hat e_1 \to \hat x$, $\hat e_2 \to \hat y$ and $\hat e_3 \to \hat z$ such that $\hat e_1 \partial_i \hat e_2 \to 0$ for $|\vec x| \to \infty$, it follows from Stokes theorem upon integrating Eq.~\eqref{eq:MerminHo}
\begin{align}
0 = Q_{\text{top},{\rm cl}} + \int_V d \vec x\, \rho_{\rm sing}.
\end{align}
In this case, there must exist singular charges that upon integration just compensate the total topological charge. This is consistent with the Poincar\'e-Hopf theorem that requires singularities in the vector fields $\hat e_{1,2}$ in case that the map $\vec x \mapsto \hat e_3$ possesses a non-zero $Q_{\text{top},{\rm cl}}$. 

\subsection{Linear spin-wave theory}

With the help of the Holstein-Primakoff representation \eqref{eq:HolsteinPrimakoff} for $\vec n$, the partition function \eqref{eq:partitionFunc} can be expressed as 
\bea
\label{eq:partitionFunc2}
\cz &=& \int \cd \psi^* \cd \psi \exp\left(-S_E[\psi^*,\psi]\right),
\eea
because the Jacobian of the transformation $\vec n(\psi^*,\psi)$ is a constant. The linear spin-wave approximation is obtained by expanding the action up to quadratic order in the magnon fields. Using the spinor notation, $\vec \Psi^T = (\psi, \psi^*)$, the result has the general form
\begin{align} \label{eq:LSWapproximation}
S^{(2)}_E[\psi^*,\psi] = \int_0^\beta d\tau \int_V d \vec x\, \Big( 
\frac{1}{2} \vec \Psi^\dagger \tau^z \partial_\tau \vec \Psi  + \frac{1}{2} \vec \Psi^\dagger H \vec \Psi
\Big),
\end{align}
where $\vec \Psi = \vec \Psi(\vec x', \tau)$ with $\vec x' = \vec x - \vec R_{\rm cl}$, and the operator matrix $H$ depends on the specific model, i.e. on $\mathcal{W}$ of Eq.~\eqref{eq:LLeq}. Importantly, after introducing the integration variable $\vec x'$, the theory \eqref{eq:LSWapproximation} for $V \to \infty$  becomes independent of the constant $\vec R_{\rm cl}$. The operator $H$ is not spatially invariant but depends on the texture positioned at $\vec x' = 0$. The theory \eqref{eq:LSWapproximation} thus constitutes a scattering problem that can be diagonalized using a bosonic Bogoliubov-deGennes transformation, for an example see Ref.~\cite{Schuette2014a}. Its stationary eigenmodes $\vec \Phi_n(\vec x')$ 
can be orthogonalized with respect to the scalar product
\begin{align}
\int d\vec x'\, \vec \Phi^\dagger_n(\vec x') \tau^z \vec \Phi_m(\vec x') \propto \delta_{nm}.
\end{align}
Here, $\tau^z$ denotes the third Pauli matrix acting on the spinor.
As the energy of the texture does not depend on the classical coordinate $\vec R_{\rm cl}$ due to the translational invariance of the system, the spin-wave problem possesses zero modes, i.e. modes with zero eigenenergy. The corresponding eigenfunctions can be derived using Eq.~\eqref{eq:HolsteinPrimakoff}. The zero modes in linear order shall be able to  compensate a small variation of $\delta \vec R_{\rm cl}$, i.e.
\begin{align}
- \partial_i \hat e_3 \delta R_{{\rm cl},i} + \sqrt{\frac{\hbar}{s}} (\hat e_- \psi^* + \hat e_+ \psi) = 0.
\label{eq:acquire_zm}
\end{align}
Considering small independent shifts $\delta \vec R_{\rm cl}$ in $x$- and $y$-direction, we can identify the corresponding unnormalized eigenfunction of the zero modes
\begin{align} \label{eq:zeromodes}
\vec \Phi_{0,x/y} =    \left( \hat e_- \partial_{x/y} \hat e_3 \atop  \hat e_+ \partial_{x/y} \hat e_3 \right).
\end{align}
Remarkably, the scalar products involving the zero modes only depend on the topological charge and not on the details of the classical texture,
\bea
\int d\vec{x}' \vec{\Phi}^{\dagger}_{0, i}\tau^z \vec{\Phi}_{0, j} &=& i4\pi Q_{{\rm top,cl}} \epsilon_{ji}.
\eea
As will be seen below, the zero modes play a central role in the discussion of the topological dipole.

\subsection{Topological charge density in the linear spin-wave approximation}
\label{sub:TopoLinSWTheory}

The Holstein-Primakoff representation \eqref{eq:HolsteinPrimakoff} can now be used to study the corrections to the classical topological charge density induced by spin-wave fluctuations. Expanding $\rt$ in the spin-wave amplitudes and after some lengthy algebra, we obtain
\begin{align} \label{eq:TopChargeSpinWave}
&\rt = \rho_{\text{top},\rm cl} + \sqrt{\frac{\hbar}{s}}\frac{i}{4\pi}\epsilon_{ij}\partial_j\left(\psi \hat e_+\partial_i \hat e_3 - \psi^* \hat e_- \partial_i \hat e_3 \right) +   
\nonumber \\&\quad
\frac{\hbar}{s} \frac{1}{8\pi} \epsilon_{ij} \partial_j 
\Big(-i \psi^* \partial_i \psi + i \psi \partial_i \psi^* + 2 (\hat e_1 \partial_i \hat e_2) \psi^* \psi\Big) + ...
\end{align}
where the dots represent corrections on the order $\mathcal{O}((\hbar/s)^{3/2})$ that we will neglect in the following. Remarkably, the lowest-order corrections possess the form of a divergence. Consequently, the total topological charge remains unchanged, i.e., $Q_{\rm top} = Q_{{\rm top},{\rm cl}}$ in case that the expressions in the parentheses of Eq.~\eqref{eq:TopChargeSpinWave}
vanish on the surface of the spatial volume, which we will assume in the following.

For the topological dipole moment, we obtain a correction to the classical limit $D_{\text{cl},i} = Q_{\rm top}R_{\text{cl},i}$,
\begin{align}
\label{eq:DipoleSpinWave}
    D_i =& D_{\text{cl}, i} + \frac{i\epsilon_{ij}}{4\pi} \sqrt{\frac{\hbar}{s}} \int_V d\vec{x} \ \vec{\Phi}^{\dagger}_{0,j} \tau^z \vec{\Psi}  + \frac{\epsilon_{ij}}{4\pi s}P_{\text{mag}, j},
\end{align}
where we made use of the zero modes \eqref{eq:zeromodes} and we introduced 
\begin{align}
\label{eq:MagnonLinearMomentum}
    P_{\text{mag},j} &= \frac{\hbar}{2}  \int_V d \vec x\,  \vec \Psi^\dagger (-i \tau^z \partial_j + \hat e_1 \partial_j \hat e_2) \vec \Psi.
\end{align}
Note that in the thermodynamic limit of an infinitely large area, $V \to \infty$, the corrections in Eq.~\eqref{eq:DipoleSpinWave} do not depend on $\vec R_{\rm cl}$ as this dependence can be absorbed by a change of integration variables, $\vec x' = \vec x - \vec R_{\rm cl}$, in both integrals.

The lowest-order correction to the dipole moment of order $1/\sqrt{s}$ in Eq.~\eqref{eq:DipoleSpinWave} is a projection of the magnon wavefunction $\vec \Psi$ onto the subspace spanned by the zero modes. This implies that, at this order, only the zero modes modify the topological dipole moment. If we consider a pure zero-mode Ansatz for the wavefunction $\psi = \sqrt{s/\hbar}\ \delta R_{\text{cl}, i} \hat e_-\partial_i\hat e_3$, the correction to $D_i(\tau)$ is simply given by $-\delta R_{\text{cl},i}(\tau)Q_{\rm top}$. It can be combined with the classical term 
$D_i(\tau) = Q_{\rm top} (R_{\text{cl}, i} - \delta R_{\text{cl}, i}(\tau))$, i.e. the coordinate effectively acquires a time dependence. 
The next-to-leading order correction involves $P_{\text{mag},j}$ defined in Eq.~\eqref{eq:MagnonLinearMomentum}.
Note that the expression for $P_{\text{mag},j} $ is manifestly invariant with respect to the $U(1)$ gauge transformation of Eq.~\eqref{eq:U(1)}. $\vec{P}_{\rm mag}$ receives contributions from both the zero modes as well as the other modes. 
Comparing Eq.~\eqref{eq:DipoleSpinWave} with Eq.~\eqref{eq:TotMomentum}, we can identify $\vec P_{{\rm mag}}$ as the $1/s$ correction to the total linear momentum. We can thus summarize for the conservation of the total linear momentum at this order of the spin-wave expansion
\begin{align} \label{eq:TotalLinearMomentum}
\partial_\tau \Big(4\pi s Q_{\rm top} \epsilon_{ji} (R_{\text{cl}, j} - \delta R_{\text{cl}, j}(\tau)) + P_{\text{mag},i}(\tau) \Big) = 0 .
\end{align}

\subsection{Quantization of the classical collective coordinate} 

The classical coordinate $\vec R_{\rm cl}$ plays a special role as it is related to the zero mode of the spin-wave problem. It is customary to treat this mode non-perturbatively by promoting it to a dynamical variable $\vec R_{\rm cl} = \vec R_{\rm cl}(\tau)$, which amounts to a quantization of $\vec R_{\rm cl}$. The degrees of freedom $\vec R_{\rm cl}$ can be introduced in the path integral of Eq.~\eqref{eq:partitionFunc2} at the expense of the zero modes of Eq.~\eqref{eq:zeromodes},
\begin{align}
\label{eq:partitionFunc2}
\cz &= \int \cd \vec R_{\rm cl} \int \cd' \psi^* \cd' \psi \exp\left(-S_E[\psi^*,\psi, \vec R_{\rm cl}]\right),
\end{align}
where the prime on $\cd' \psi^* \cd' \psi$ indicates that the path-integration excludes the zero modes. Technically, this can be achieved using the Fadeev-Popov technique as explained in Appendix \ref{app:FadeevPopov}.

The linear spin-wave approximation of the action in the presence of a time dependent $\vec R_{\rm cl}$ requires more care. The dynamical Berry phase now generates an additional term, and we get instead of  Eq.~\eqref{eq:LSWapproximation} \cite{Schuette2014a,Stone1996},
\begin{align} \label{eq:LSWapproximation2}
&S^{(2)}_E[\psi^*,\psi, \vec R_{\rm cl}] = S^{(2)}_E[\psi^*,\psi] +
\\ \nonumber 
& + 
 \int_0^\beta d\tau\, \Big[2\pi s Q_{\rm top}\epsilon_{ij}   R_{{\rm cl},j} - P'_{{\rm mag}, i} \Big] \frac{i}{\hbar} \partial_\tau  R_{{\rm cl},i} ,
\end{align}
where $\vec P'_{\rm mag}$ depends on the magnon wavefunction and it has the same form as Eq.~\eqref{eq:MagnonLinearMomentum}. The prime on $\vec P'_{\rm mag}$ is a reminder that it does not include the magnon zero modes as they are projected out by the path integral \eqref{eq:partitionFunc2}. Note that neither the first term in Eq.~\eqref{eq:LSWapproximation2} nor $\vec P'_{\rm mag}$ depends on $\vec R_{\rm cl}$ in the thermodynamic limit. Variation of the action \eqref{eq:LSWapproximation2} with respect to the classical coordinate thus yields the conservation law
\begin{align} \label{eq:ConservationLawCollectiveCoordinate}
\partial_\tau \Big( 4\pi s Q_{\rm top}  \epsilon_{ji} R_{{\rm cl},j}(\tau) + P'_{{\rm mag}, i}(\tau) \Big) = 0 .
\end{align}
This just describes the conservation of total linear momentum which should be compared to Eq.~\eqref{eq:TotalLinearMomentum}.
Due to the quantization of the collective coordinate, $R_{{\rm cl},j}(\tau)$ is now fully dynamic at the expense of the second term $P'_{{\rm mag}, i}(\tau)$ that excludes the zero modes. Importantly, the dynamics of $R_{{\rm cl},j}$ and $P'_{{\rm mag}, i}$ are coupled in a non-trivial manner because the total linear momentum, i.e., the topological dipole moment is conserved.

\subsection{Effective action for the collective coordinate}
\label{sub:EfActionCollCoord}

An effective action for the collective coordinate can be derived by integrating out the magnon modes. The effective action for the classical collective coordinate $R_{{\rm cl},j}(\tau)$ can only be derived perturbatively as this coordinate is interacting with the magnon modes. Perturbatively integrating out the magnon modes in lowest order, we obtain 
\begin{align} \label{eq:EffectiveAction1}
&S_{E,{\rm eff}}[\vec R_{\rm cl}] \approx \int_0^\beta d\tau\, 2\pi sQ_{\rm top}  \epsilon_{kj}  R_{{\rm cl},j} \frac{i}{\hbar} \partial_\tau  R_{{\rm cl},k} +
\\\nonumber
&\!\int_0^\beta d\tau d\tau' \frac{1}{2\hbar^2} \partial_\tau' R_{{\rm cl},i}(\tau') \langle P'_{{\rm mag}, i}(\tau')P'_{{\rm mag}, j}(\tau)\rangle \partial_\tau R_{{\rm cl},j}(\tau) ,
\end{align}
where the expectation value in the second line is evaluated with respect to the path-integration over the magnon fields in Eq.~\eqref{eq:partitionFunc2}. For the specific model of a chiral magnet, it was argued in Ref.~\cite{Psaroudaki2017} that this expectation value reduces in the low-energy limit to an effective mass $m_{\rm eff}$,
\begin{align} 
\langle P'_{{\rm mag}, i}(\tau')P'_{{\rm mag}, j}(\tau)\rangle \approx m_{\rm eff} \delta(\tau - \tau'),
\end{align}
such that the classical coordinate acquires inertia. 

Alternatively, we might exploit the conservation law of Eq.~\eqref{eq:ConservationLawCollectiveCoordinate} and perform a change of integration variables
\begin{align} \label{eq:DecompCoord}
R_i = R_{{\rm cl},i} + \frac{1}{4 \pi s Q_{\rm top}} \epsilon_{ij}  P'_{{\rm mag},j},
\end{align}
where $D_i = R_i Q_{\rm top}$ is the topological dipole moment renormalized by fluctuations. We emphasize that Eq.~\eqref{eq:DecompCoord} is just the linear spin wave representation of Eq.~\eqref{eq:collectivecoor}.
The partition function then becomes in the linear spin-wave approximation,
\begin{align}
\label{eq:partitionFunc3}
\cz^{(2)} &= \int \cd \vec R \int \cd' \psi^* \cd' \psi \exp\left(-S^{(2)}_E[\psi^*,\psi, \vec R]\right),
\end{align}
where the action is obtained with the help of Eq.~\eqref{eq:LSWapproximation2},
\begin{align} \label{eq:TheoryFactorization}
&S^{(2)}_E[\psi^*,\psi, \vec R] =  \int_0^\beta d\tau\, 
2\pi s Q_{\rm top} \epsilon_{kj} R_{j} \frac{i}{\hbar} \partial_\tau  R_{k} 
\\ \nonumber 
& +  S^{(2)}_E[\psi^*,\psi] +
 \int_0^\beta d\tau\, 
 \epsilon_{kj} \frac{1}{8\pi sQ_{\rm top}}  P'_{{\rm mag}, j}  \frac{i}{\hbar} \partial_\tau  P'_{{\rm mag}, k}.
 \end{align}
The magnon modes completely decouple from the topological dipole moment and, as a result, the partition function factorizes. Consequently, the effective action for $\vec R$ is, at this linear order of spin-wave approximation, exactly given by 
\begin{align} \label{eq:EffectiveAction2}
&S_{E,{\rm eff}}[\vec R ] = \int_0^\beta d\tau\,  2\pi s Q_{\rm top} \epsilon_{kj} R_{j} \frac{i}{\hbar} \partial_\tau  R_{k} .
\end{align}
Importantly, this action for $R_i = D_i/Q_{\rm top}$ is consistent with the commutator for the dipole operator $\hat D_i$ of Eq.~\eqref{eq:Dcommutator}.

\section{Discussion}
\label{sec:Discussion}

In this work, we demonstrated that the topological dipole $D_i$ of a two-dimensional spin texture is a conserved quantity and independent of time for a continuum quantum field theory with translational invariance provided that hedgehog defects in 2+1 spacetime are energetically suppressed and absent. The underlying reason is that the topological dipole moment $D_i$ is related to the linear momentum $P_i$, and $P_i$ remains conserved for appropriate boundary conditions in the presence of thermal and quantum fluctuations of the magnetization. This generalizes the classical considerations of previous work by Papanicolaou and Tomaras \cite{Papanicolaou1991}. Our findings have various implications that we discuss in the following.

\subsection{Skyrmion mass problem for the topological dipole coordinate}
For a magnetic skyrmion texture with a non-zero topological charge $Q_{\rm top} \neq 0$, the dipole can be interpreted as a collective coordinate $R_i = D_i/Q_{\rm top}$ identifying the position of the skyrmion. We demonstrated that this coordinate obeys the undamped Thiele equation of Eq.~\eqref{eq:Pconservation}, i.e., the skyrmion does not move in the absence of external forces $\partial_t \vec R = 0$. This implies that fluctuations of the magnetization do not self-generate any corrections neither damping nor inertia. In particular, the skyrmion mass for the topological dipole coordinate vanishes. As will be clear in the next subsections, this result not only simplifies the dynamical description, it also yields a profound connection between magnetic skyrmions, fractons and the physics of the lowest Landau level.

\begin{figure}[t]
    \centering
    \includegraphics[width=0.5\linewidth]{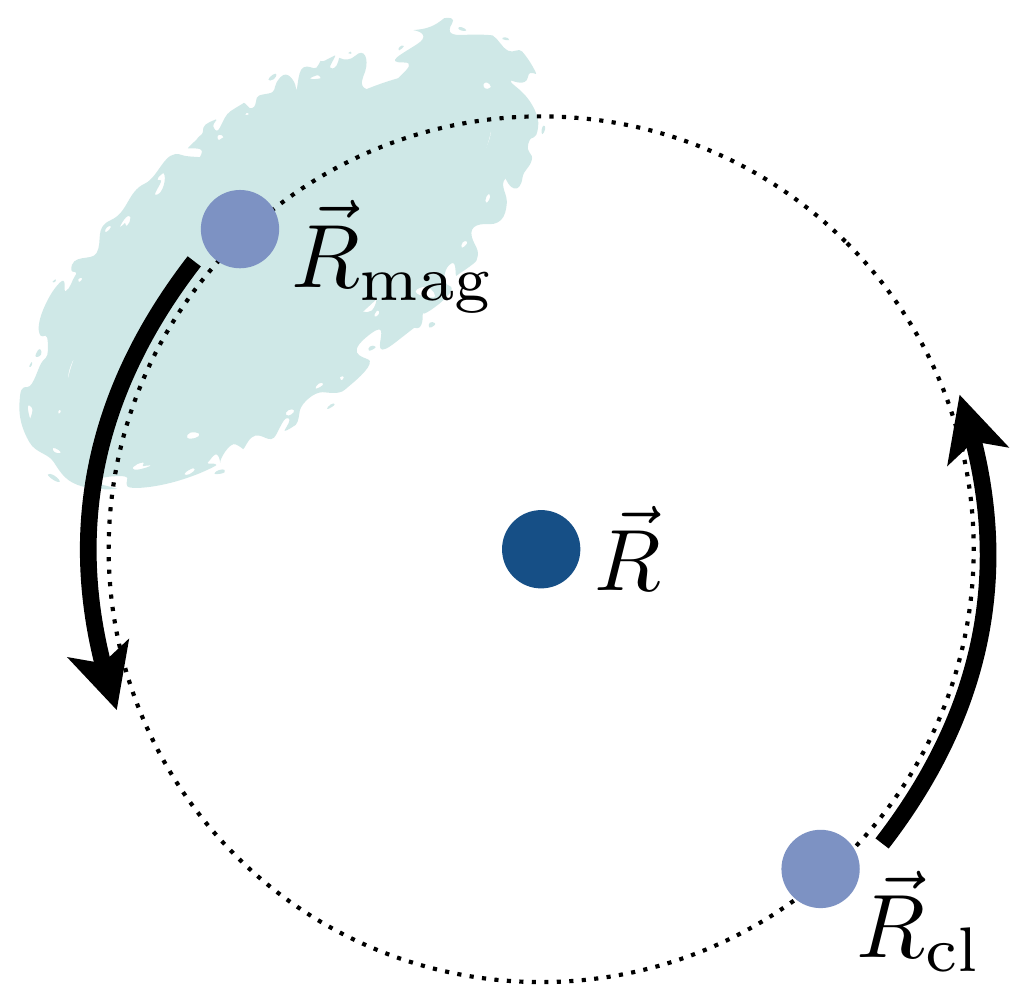}
    \caption{Schematic illustration of the collective coordinate $R_i = D_i/Q_{\rm top}$ of a topological spin texture with the topological charge $Q_{\rm top} \neq 0$ and the topological dipole $D_i$, that decomposes into a classical and a magnon component, $R_i = R_{\text{cl},i} + R_{\text{mag},i}$. Whereas the dipole moment is conserved, $\partial_t R_i = 0$, the magnon excitations impose a non-trivial dynamics on the classical coordinate such that the time-evolution of $R_{\text{cl},i} $ is counterbalanced by a magnon cloud with coordinate $R_{\text{mag},i}$.
    }
    \label{fig:cloud}
\end{figure}

Our explicit  treatment of spin-wave fluctuations in Section \ref{sec:Spinwavetheory} transparently shows how this can be reconciled with previous calculations of Psaroudaki {\it et al.} \cite{Psaroudaki2017} who found a finite skyrmion mass $m_{\rm eff} \neq 0$. 
It is important to realize that
the collective coordinate, $R_i = R_{{\rm cl},i} + R_{{\rm mag},i}$ see Eq.~\eqref{eq:DecompCoord}, decomposes into a classical part $R_{{\rm cl},i}$ representing the translational magnon zero-mode and a contribution attributed to the remaining magnon excitations, $R_{{\rm mag},i} = \frac{1}{4\pi s Q_{\rm top}} \epsilon_{ij} P'_{{\rm mag},j}$, where $P'_{{\rm mag},j}$ is the associated linear momentum. As the topological dipole moment is conserved, the collective coordinate is independent of time, $\partial_t R_i = 0$. This implies that the magnon excitations impose a non-trivial dynamics on the classical component because $\partial_t R_{{\rm cl},i} = - \partial_t R_{{\rm mag},i}$. Consequently, after integrating out the spin-wave excitations, the equation of motion for $R_{{\rm cl},i}$ is renormalized and this renormalization was calculated in Ref.~\cite{Psaroudaki2017}. This means that the coordinate in Eq.~\eqref{eq:withmass} is distinct from the coordinate of Eq.~\eqref{eq:Pconservation}, as the former should be rather identified with $R_{{\rm cl},i}$. The finite mass $m_{\rm eff}$ allows a cyclotron motion for the classical component $R_{{\rm cl},i}$ with frequency $\omega_c = |\vec G|/m_{\rm eff}$, where $|\vec G| = 4\pi s |Q_{\rm top}|$, that revolves around a constant $R_i$. This revolving motion must be counterbalanced by the motion of a magnon cloud represented by $R_{{\rm mag},i}$, see the illustration in Fig.~\ref{fig:cloud}. 

Could this decomposition of the internal dynamics with frequency $\omega_c$ be detected experimentally? One of our major finding is that the topological dipole degree of freedom, $R_i$,  factorizes in linear spin wave theory and, as a result, a specific magnon-magnon interaction is generated, see the last term in Eq.~\eqref{eq:TheoryFactorization}. The internal dynamics should be fully encoded in the impact of this generated non-linearity. Its influence will, however, depend on the details of the magnetic system. It could generate in the dynamical spin structure factor either a well-defined magnon-magnon bound state with frequency $\omega_c$ or rather an undetectable overdamped feature. The study of this magnon-magnon interaction in various magnetic systems is an interesting avenue for future research.

\subsection{Lowest Landau level correspondence} 

It was already pointed out in Ref.~\cite{Papanicolaou1991} that the algebra for the topological dipole moment $D_i$ of a spin texture given by the Poisson bracket \eqref{eq:PoissonTopDip} is related to that of the guiding center of an electron in a magnetic field.
The Poisson bracket for $D_i$ also served as a starting point for the canonical quantization of skyrmions discussed in Ref.~\cite{Ochoa2019}. Using the path-integral formalism in Section \ref{sec:quantum}, we confirmed that it indeed generalizes to the quantum theory, see Eq.~\eqref{eq:Dcommutator}. For a texture with a non-zero $Q_{\rm top} < 0$, the commutator for the quantized collective coordinate is $[\hat R_i, \hat R_j] = i \ell^2 \epsilon_{ij}$, where $\ell = \sqrt{\frac{\hbar}{4\pi s |Q_{\rm top}|}}$ is identified with the magnetic length of the corresponding electron problem. This commutator is also consistent with the effective action for the coordinate given in Eq.~\eqref{eq:EffectiveAction2}. 
Quantum effects are expected to become important if the length $\ell$ becomes appreciably large compared with intrinsic lengths of the magnetic system, e.g. the skyrmion radius $\ell_s$, namely $\ell/\ell_s \sim 1$.

Reference~\cite{Ochoa2019} discussed the consequences of non-commuting position operators $\hat R_i$ for quantum skyrmions
and its relation to quantum Hall physics. Due to the non-trivial commutator for the components of $\hat R_i$, the band structure of quantum skyrmions will be topologically non-trivial with bands possessing finite Chern numbers. This implies the presence of chiral edge modes similar to the integer quantum Hall effect. In the latter context, these chiral edge modes can be semiclassically interpreted as skipping orbits of electrons such that their guiding center is propagating along the edge in a direction singled out by the Lorentz force.  

We found that there is an even closer correspondence with the physics of the lowest Landau level. Similar to the charge density projected to the lowest Landau level, the topological density $\rt$ as a generator of area-preserving diffeomorphism satisfies the Girvin-MacDonald-Platzman algebra \cite{Girvin1986} as pointed out previously in Ref.~\cite{Du2022}. On the classical level, this follows straightforwardly from the definition of the Poisson bracket for the magnetization field, see Eq.~\eqref{eq:PoissonTopCharge}, but we found that it also generalizes to the quantum theory, see Eq.~\eqref{eq:GirvinMacDonaldPlatzman}. The magnetic system and the lowest Landau level thus share similar properties that derive from this algebra. For example, the Fourier transform of the static structure factor, $s(\vec r) = \langle \rt(\vec r) \rt(0)\rangle - \langle \rt\rangle^2$, of a skyrmion liquid is expected to vanish in the long-wave length limit as $s(\vec k) \sim |\vec k|^4$ \cite{Girvin1986}. Moreover, the Goldstone mode of the skyrmion crystal possesses a dispersion $\omega \sim \vec k^2$ for small wavevectors  \cite{Zang2011, Oleg2011} similar to the phonon of the Wigner crystal at small electron filling fractions \cite{Du2022}. This analogy suggests the exciting prospect that skyrmion liquids might be classified in a similar manner as fractional quantum Hall states, which deserves further investigations. 

\subsection{Skyrmion-fracton correspondence} 

A skyrmion with a finite charge $Q_{\rm top} \neq 0$ cannot move and stays immobile due to the conservation of the topological dipole
$\partial_t D_i = Q_{\rm top} \partial_t R_i = 0$.
In contrast, a spin texture with vanishing net charge $Q_{\rm top} = 0$, e.g., a topological charge dipole consisting of a skyrmion-antiskyrmion pair with individual charges $\pm 1$, can roam around. Indeed, solutions of the Landau-Lifshitz equation corresponding to mobile topological charge dipole structures have already been discussed in easy-axis chiral magnets \cite{Komineas2015}.
The conservation of the topological dipole is rooted in the continuity equation \eqref{eq:continuity2} for the topological charge density $\rt$ where the topological current itself is given by a divergence of a rank-two tensorial object, see Eq.~\eqref{eq:momentumConservation}.

Strikingly, this is analogous to the dynamical properties of fractons in scalar charge gauge theory \cite{Pretko1, Pretko2}. In such a theory, a $U(1)$ charge and its first moment are conserved, which manifests itself in a continuity equation of the form as in Eq.~\eqref{eq:continuity2}. As a result, fractons carrying a $U(1)$ charge are immobile whereas dipoles are able to move. This analogy suggests that magnetic skyrmions are characterized by fractonic behavior and provide a platform for studying the associated exotic phenomena. These include, but are not limited to, $(i)$ anomalous hydrodynamics of a fracton fluid which does not obey Fick's law for a diffusion process but instead exhibits a subdiffusive behaviour \cite{Morningstar2020, Gromov2020, Zhang2020, Feldmeier2020}, $(ii)$ ergodicity-breaking physics giving rise to interesting thermalization processes \cite{Pai2019, Khemani2020, Sala2020, PretkoRev}, and $(iii)$ an emergence of high-rank tensor gauge fields which are compatible with the conservation of higher moments \cite{Pretko2, Pretko2018gauge, Du2022}.

In contrast to the original proposal where the conserved fracton charge generates an internal $U(1)$ transformation \cite{Pretko2, Pretko2018gauge}, the topological charge density $\rt$ generates area-preserving diffeomorphisms, see Eq.~\eqref{eq:generatorAPD}, i.e., spatial translational symmetry transformations. As elaborated on in Ref.~\cite{Du2022}, this has consequence for the structure of the emerging gauge theory. The corresponding gauge potential in the present case will be related to the spatial metric and the associated gauge theory might be interpreted as a theory of linearized gravity. This is again similar to the physics of the lowest Landau level. 

The relation between translation symmetry and the fractonic properties of magnetic skyrmions will be advantageous for its experimental exploration. There are many accessible ways to fully or partially break the translational symmetry, e.g. via a magnetic field gradient, an applied current, an atomic lattice \cite{Balents2016} or an engineered potential \cite{Mueller2015,Psaroudaki2020}. This provides a promising freedom to tune the magnetic skyrmion system into and out of the fractonic regime and to explore the physics in various symmetry-broken phases.

\section{Acknowledgement}
We would like to thank Volodymyr Kravchuk and Ho Tat Lam for very helpful discussions. M.G. acknowledges support from Deutsche Forschungsgemeinschaft (DFG) via Project-id 403030645 (SPP 2137 Skyrmionics) and Project-id 445312953.

\appendix

\section{Fadeev-Popov technique}
\label{app:FadeevPopov}

In order to put the partition function into the form of Eq.~\eqref{eq:partitionFunc2}, we employed the Faddeev-Popov technique \cite{Gervais1975,Braun2012} by inserting the following identity into the path integral
\begin{align}
\label{eq:FPtechnique}
\int \cd \vec R_{\rm cl}\ \delta[\Delta_1(\vec R_{{\rm cl}})]\delta[\Delta_2(\vec R_{{\rm cl}})]  
\Big|\det\Big(\frac{\partial \Delta_i}{\partial R_{{\rm cl},j}}\Big)\Big| 
= 1,
\end{align}
with a suitable choice for the two functions $\Delta_i$ with $i = 1,2$; for example, 
\begin{align}
\Delta_i(\vec R_{{\rm cl}}) &= \int_V d\vec x' \vec n(\vec x' + \vec R_{{\rm cl}}, \tau) (\vec n_{\rm cl}(\vec x') \times \partial_i \vec n_{\rm cl}(\vec x')),
\\
\frac{\partial \Delta_i}{\partial \vec R_{{\rm cl},j}} &= \int_V d\vec x' \partial_j \vec n(\vec x' + \vec R_{{\rm cl}}, \tau) (\vec n_{\rm cl}(\vec x') \times \partial_i \vec n_{\rm cl}(\vec x')).
\end{align}
The merit of this choice becomes apparent when evaluating it in lowest order in the spin-wave approximation. Inserting for $\vec n$ the Holstein-Primakoff representation \eqref{eq:HolsteinPrimakoff} and expanding $\Delta_i$ up to linear order we get 
\begin{align}
\Delta_i(\vec R_{{\rm cl}}(\tau)) \approx -i \sqrt{\frac{\gamma \hbar}{M_s}} \int_V d\vec x'  \vec \Psi^\dagger(\vec x',\tau) \tau^z \vec \Phi_{0,i}(\vec x').
\end{align}
The function $\Delta_i$ projects for each time $\tau$ the spin-wave function onto the zero modes of Eq.~\eqref{eq:zeromodes} such that the delta functions in Eq.~\eqref{eq:FPtechnique} suppresses their amplitudes. The matrix of the Jacobian can be also evaluated and we get in zeroth order,
\begin{align}
\frac{\partial \Delta_i}{\partial \vec R_{{\rm cl},j}} \approx \int_V d\vec x' \partial_j \hat e_3(\vec x') (\vec e_3(\vec x') \times \partial_i \hat e_3(\vec x')) = 4\pi \epsilon_{ij} Q_{\rm top, cl}.
\end{align}
Absorbing constant factors into the measure of the path integral we thus arrive at Eq.~\eqref{eq:partitionFunc2}.

\bibliography{refs}

\end{document}